\documentclass[11pt]{article}

\usepackage[a4paper,margin=2.5cm]{geometry}
\usepackage{authblk}
\usepackage{comment}
\usepackage{amsmath}
\usepackage{amssymb}
\usepackage{amsthm}
\usepackage{booktabs}
\usepackage{multirow}
\usepackage{colortbl}
\usepackage{xcolor}
\usepackage{graphicx}
\usepackage[round,authoryear]{natbib}
\usepackage{xurl}
\usepackage[hidelinks]{hyperref}

\hypersetup{
  pdftitle={Testing similarity of competing risks models by comparing transition probabilities},
  pdfauthor={Zoe Kristin Lange, Maryam Farhadizadeh, Holger Dette, and Nadine Binder}
}

\newtheorem{theorem}{Theorem}
\newtheorem{remark}{Remark}
\newcommand{\uproman}[1]{\uppercase\expandafter{\romannumeral#1}}

\makeatletter
\newcounter{algorithm}
\newcommand{\fname@algorithm}{Algorithm}
\newenvironment{breakablealgorithm}
{
  \par\noindent
  \refstepcounter{algorithm}
  \hrule height.8pt depth0pt \kern2pt
  \centering
  \renewcommand{\caption}[2][\relax]{
    {\raggedright\textbf{\fname@algorithm~\thealgorithm} ##2\par}
    \ifx\relax##1\relax
      \addcontentsline{loa}{algorithm}{
        \protect\numberline{\thealgorithm}##2}
    \else
      \addcontentsline{loa}{algorithm}{
        \protect\numberline{\thealgorithm}##1}
    \fi
    \kern2pt\hrule\kern2pt
  }
}
{
  \par\kern2pt\hrule\relax
}
\makeatother

\newcommand{\keywords}[1]{%
  \par\medskip\noindent\textbf{Keywords:} #1\par\medskip
}

\title{Testing similarity of competing risks models by comparing transition probabilities}

\author[1]{Zoe Kristin Lange\thanks{Corresponding author: zoe.lange@ruhr-uni-bochum.de}}
\author[2,3]{Maryam Farhadizadeh}
\author[1]{Holger Dette}
\author[2,3]{Nadine Binder}

\affil[1]{Department of Mathematics, Ruhr University Bochum, Bochum, Germany}
\affil[2]{Institute of General Practice/Family Medicine, Medical Center and Faculty of Medicine, University of Freiburg, Freiburg, Germany}
\affil[3]{Center for Data Analysis, Modeling and AI, University of Freiburg, Freiburg, Germany}

\date{}
\begin{document}

\maketitle

\begin{abstract}
Assessing whether patient populations exhibit comparable event dynamics is important for evaluating treatment equivalence, pooling cohorts and comparing clinical pathways. Existing similarity tests for competing risks models measure distances between transition intensities, which describe instantaneous event rates. In biomedical applications, similarity may be more naturally formulated through transition probabilities, which quantify cumulative event risks over a clinically relevant horizon. Assuming constant cause-specific transition intensities, we develop a framework for testing similarity based on a maximum-type distance between vectors of transition-probability functions. We propose a constrained parametric bootstrap test and establish asymptotic level control and consistency under administrative and independent exponential random right censoring. The constant-intensity formulation is motivated by small-data settings in which few events are observed and nonparametric estimators may be unstable. Simulations across sample sizes, censoring mechanisms and degrees of dissimilarity show that the proposed test can attain larger finite-sample rejection probabilities than an intensity-based benchmark under comparable alternatives. An application to routine prostate cancer data illustrates how the procedure identifies the smallest examined margin for which similarity of 90-day readmission-probability functions can be established under the fitted model. The method provides an interpretable and practically implementable basis for similarity assessment in parametric competing risks models.

\end{abstract}


\section{Introduction}

Understanding the clinical events experienced by patient populations over time is a key objective of medical and health services research. Routinely collected hospital data increasingly allow such event dynamics to be studied over clinically relevant follow-up periods initiated by, for example, a hospital admission, diagnostic procedure, or treatment. In many applications, these dynamics are naturally described by competing risks models, where patients start in a common initial state and may transition to one of several mutually exclusive event states, such as different treatments, complications, readmissions, or other clinical outcomes \citep{AndersenCountingProcesses,AndersenKeiding,Beyersmann,Putter}.

A central question is whether these event dynamics are comparable across patient groups defined by different diagnostic pathways, treatment strategies, or institutional settings. Comparing the probabilities of clinically relevant events provides an interpretable means of assessing the similarity of the corresponding patient pathways. Establishing such similarities facilitates clinical interpretation and may also support combining cohorts in subsequent statistical analyses \citep{Binder}.

Within this setting, assessing whether two groups of patients share similar transition behavior becomes a problem of comparing two stochastic processes. However, formal statistical methods for testing {\it similarity} between such processes remain scarce, even though these comparisons are fundamental for evaluating equivalence between treatment strategies, care pathways, or healthcare systems. While competing risks analyses do offer a number of established hypothesis tests designed to detect significant {\it differences} between groups \citep[see, for example][and the references therein]{lyu_comparison_2020, sestelo_method_2024}, these methods are not designed to establish whether differences are sufficiently small to regard the processes as similar. On the other hand, similarity testing has found considerable interest in other applications such as equivalence testing of pharmacokinetic and pharmacodynamic models \citep{kaneko2023method}, 
the comparison of dose response profiles \citep{kxaa058} and the assessment of consistency in clinical trials \citep{grill2020assessing}. Most of these approaches were inspired by equivalence testing concepts from pharmacokinetics, where bioequivalence between two drug concentration profiles is determined based on minimal differences \citep{Hauschke}.

In recent years, methodological developments have introduced formal approaches for testing the similarity between competing risk models.
\citet{Binder} proposed a framework for comparing competing risk models based on cause-specific transition intensities. In this framework, each patient group is modeled as a continuous-time Markov process with constant hazards. 
This parametric formulation is motivated by small data settings, where the number of observed events and the total time at risk may be limited and nonparametric approaches typically can be less efficient or unstable \citep[see e.g. Section 3.1 in][]{jullum_what_2019}.
Specifically, the similarity test of \citet{Binder} evaluates whether the transition intensities of two groups differ by more than a predefined threshold. Their testing procedure is not based on the intersection-union principle as the classic two one-sided test (TOST) in bio-equivalence studies, but involves a bootstrap approach, which mimics the (asymptotically) uniformly powerful test \citep[see][]{romano}.
\citet{Moellenhoff} extended this approach to parametric intensity models beyond the case of constant hazard values, including Gompertz and Weibull distributions. Additionally, they standardized comparisons across competing risks using a global maximum-type test statistic. This improves the informative power and interpretability compared to separate tests for each cause. These studies established the first systematic methods for testing similarity in event history models and demonstrated that intensity-based bootstrap tests can effectively control type I error and achieve high power in many scenarios.

The present paper takes these developments as a starting point and investigates whether similarity testing can be formulated directly in terms of transition probabilities rather than transition intensities. This shift is motivated by the distinct scientific interpretations of transition intensities and transition probabilities. Cause-specific transition intensities describe instantaneous event rates among patients who remain in the initial state and are therefore informative about the local transition mechanisms. By contrast, transition probabilities quantify the cumulative probability of experiencing a particular event over a clinically relevant follow-up period and thus directly address questions such as the probability of readmission within 30 or 90 days. Moreover, the transition probability of a given event depends not only on its own cause-specific intensity but also on the intensities of all competing events. An increase in the intensity of one event may therefore reduce the probability of another event, even if the latter event's own intensity remains unchanged. Consequently, differences between individual transition intensities do not translate directly into differences in cumulative event risks. Modest but persistent differences in intensities may accumulate into clinically relevant probability differences, whereas larger instantaneous differences may have only limited consequences over the follow-up period. Comparing transition-probability functions therefore aligns the similarity criterion with the cumulative risks experienced by patients. In particular, using a maximum-type distance allows similarity to be defined relative to a prespecified threshold representing the largest clinically acceptable absolute difference in event probabilities across event types and follow-up time. \citep{AndersenKeiding,Geskus}.
At the same time, moving from transition intensities to transition probabilities is technically nontrivial. Transition probabilities are nonlinear functions of the underlying transition intensities, and maximum-type distances between transition probability functions lead to a more involved asymptotic analysis as well as more elaborate constrained optimization problems. 

In this paper, we introduce a novel similarity testing framework for competing risks models that focuses on transition probabilities rather than transition intensities. We define a maximum-type distance between the transition probability functions of two competing risks processes and construct a constrained parametric bootstrap test for the null hypothesis of model dissimilarity. The paper is structured as follows. In Section \ref{Section2}, we define the modeling setting and outline the algorithmic procedure for testing the global hypotheses. We state that the resulting test has asymptotic level $\alpha$ and is consistent under administrative and exponential random right censoring while the respective proof can be found in Section~\ref{appendix} of the Supplementary Material. In Section \ref{Chapter Simulation}, we show through simulations that the procedure has good finite-sample calibration and can yield larger rejection probabilities than the previously proposed benchmark based on transition intensities \citep{Moellenhoff} under comparable alternatives. In Section \ref{Section 4}, we apply the method to routine clinical data on hospital readmission after radical prostatectomy, illustrating how the probability-based formulation leads to an interpretable similarity threshold on the scale of event probabilities. Finally, we conclude with a general discussion.

\section{Modeling setting and similarity test approach} \label{Section2}

\subsection{Competing Risk Models and Estimation of Transition Probabilities} \label{Section2.1}

For modeling the event dynamics observed for two different groups of patients, we follow \citet{AndersenCountingProcesses} and consider two independent Markov processes in continuous time
\begin{equation*}
   X^{(1)}=  (X^{(1)}(t))_{t \geq 0} ~,~ X^{(2)} = (X^{(2)}(t))_{t \geq 0}
\end{equation*}
with finite state space $\{0,1,\dots,k\}$ (for a fixed $k \in \mathbb{N}$).
State $0$ corresponds to the starting state for every patient, that is 
$
    \mathbb{P}(X^{(\ell)}(0)=0)=1, 
$
while we identify the absorbing competing risks with the other states $1,\dots,k$. The probability for a transition of a patient from state
$0$ at time $s$ to one of the other states $j \in \{1,2,\dots,k\}$ at time $t > s \geq 0 $ is given by
\begin{equation*}
    P^{(\ell)}_{0j}(s,t) = \mathbb{P}(X^{(\ell)}(t)=j \vert X^{(\ell)}(s)=0),
\end{equation*}
and the cause-specific transition intensities from state $0$ to state $j$ of the process $X^{(\ell)}$  are defined by
  $ \alpha^{(\ell)}_{0j}(t) = \lim_{\Delta t \rightarrow 0} \frac{P^{(\ell)}_{0j}(t,t+\Delta t)}{\Delta t}.$
Both the transition intensities
$\alpha^{(\ell)} = \big \{ (\alpha_{0j}^{(\ell)}(t))_{j=1,\dots,k} | t \geq 0  \big \} $
as well as the transition probabilities 
$    P^{(\ell)} = \big\{(P_{0j}^{(\ell)}(s,t))_{j=1,\dots,k} | t > s \geq 0 \big\}
$
characterize the  competing risks model $X^{(\ell)}$ and can therefore be used for measuring the  similarity  between $X^{(1)}$ and $X^{(2)}$.  
In contrast to \citet{Binder} and \citet{Moellenhoff}, who discussed the similarity between competing risk models using measures based on transition intensities,
we focus in this article on similarity measures based on transition probabilities. This choice is motivated by the fact that transition probabilities are often closer to the scientific question in biomedical applications.
 
Our methodological approach is motivated by a concrete application involving routinely collected clinical data from prostate cancer patients treated by radical prostatectomy. In this application, only a relatively small number of transitions involving certain states are observed, which makes a  nonparametric estimation of the transition intensities difficult. 
 Although the proposed methodology can, in principle, be applied to general parametric models for the transition intensities, we focus throughout the paper on models with constant transition intensities. This allows us to simplify the notation substantially and to develop a rigorous theoretical framework, in particular, to establish formal statistical guarantees for the proposed procedure.
 
The constant-intensity model has also been considered by \citet{Moellenhoff}. To facilitate comparison with their work, we investigate similar censoring mechanisms. Specifically, we consider administrative censoring at a fixed time ($T>0$) in this section and random right censoring in Section~\ref{SectionRandomRightCensoring}.

Under the assumption that all transition intensities are constant, the Markov processes are time-homogeneous and the transition probabilities depend only on the time difference $t-s$. Taking $s=0$, we obtain
\begin{align} \label{Eq:ProbabilityFromIntensity}
    P_{0j}^{(\ell)}(t) &= \mathbb{P}(X^{(\ell)}(t)=j \vert X^{(\ell)}(0)=0) =
    - \frac{(-1+e^{-\alpha^{(\ell)}_{0} t})\alpha^{(\ell)}_{0j}}{\alpha^{(\ell)}_{0}}
\end{align}
for $j=1,\dots,k$,  where $
    \alpha^{(\ell)}_{0} = \sum_{j=1}^{k} \alpha^{(\ell)}_{0j}
$
is the all-cause hazard and the second  equality  in  \eqref{Eq:ProbabilityFromIntensity} follows from \citet{Albert}.
We estimate the transition intensities by the maximum likelihood method. By \citet{AndersenKeiding}, the distribution of the survival time $T^{(\ell)}$ of individuals in group $\ell =1,2$ is specified by $S^{(\ell)}(t) := \mathbb{P}(T^{(\ell)} \geq t) = \exp\left(- \sum_{j=1}^k \alpha_{0j}^{(\ell)} t\right)$.
For sample sizes $n_1,n_2 \in \mathbb{N}$, let $X^{(1)}_1,\dots,X^{(1)}_{n_1}$ and $X^{(2)}_1,\dots,X^{(2)}_{n_2}$ be iid samples of the processes $X^{(1)}$ and $X^{(2)}$ with survival times $T^{(1)}_1,\dots,T^{(1)}_{n_1}$ and $T^{(2)}_1,\dots,T^{(2)}_{n_2}$, respectively. They represent the healthcare pathways of two groups of patients. In the case of administrative censoring at time point $T \in \mathbb{R}_{>0}$, we observe the survival time $\tilde{T}_i^{(\ell)} = \min\{T_{i}^{(\ell)}, T\}$ and the state $X_i^{(\ell)}(\tilde{T}_i^{(\ell)})$ for each sample $X_i^{(\ell)}$.
Then, as in \citet{AndersenCountingProcesses}, the log-likelihood function, based on the observations of group $\ell$ with administrative censoring, is given by
\begin{align} \label{Eq:LikelihoodTypeOneCensoring}
    \log \mathcal{L}^{(\ell)}(\alpha^{(\ell)})
    ={}& \sum_{i = 1}^{n_\ell} \log(S^{(\ell)}(\tilde{T}_i^{(\ell)})) + \sum_{i = 1}^{n_\ell} \sum_{j = 1}^{k}
    I\{X_i^{(\ell)}(\tilde{T}_i^{(\ell)}) = j\}
    \log(\alpha_{0j}^{(\ell)}).
\end{align}
Maximizing the log-likelihood function yields the maximum likelihood estimators (MLEs)
\begin{equation} \label{Eq:MLEIntensities}
    \hat{\alpha}^{(\ell)}_{0j} = \frac{ \sum_{i=1}^{n_\ell} I\{X_i^{(\ell)}(\tilde{T}_i^{(\ell)}) = j\}}{\sum_{i=1}^{n_\ell} \tilde{T}_i^{(\ell)}}
\end{equation}
for the transition intensities $\alpha_{0j}^{(\ell)}$ with $j = 1,\dots,k$, $\ell =1,2$.
In other words, we estimate the transition intensities by dividing the number of individuals in group $\ell$ that transitioned from state $0$ to state $j$ by the total survival time of all individuals in group $\ell$.

Estimators of the transition probability functions are given by $\{ \hat{P}^{(\ell)}_{0j}(t) \vert t \in [0,T] \}$ where $\hat{P}^{(\ell)}_{0j}(t)$
is defined by \eqref{Eq:ProbabilityFromIntensity} with
${\alpha}^{(\ell)}=\hat{\alpha}^{(\ell)}$
for $j=1,\dots,k$.

\subsection{Random Right Censoring} \label{SectionRandomRightCensoring}
According to \citet{AndersenCountingProcesses}, we model random right
censoring by introducing, for each group $\ell=1,2$, a random censoring
time $C^{(\ell)}$ that is independent of $X^{(\ell)}$. We assume that
$C^{(\ell)}$ has the parametric distribution function
$G^{(\ell)}\left(t,\psi^{(\ell)}\right)$ with density function
$g^{(\ell)}\left(t,\psi^{(\ell)}\right)$, where $\psi^{(\ell)}$ denotes
the corresponding distribution parameter.
For each group $\ell=1,2$, let $(X_i^{(\ell)},C_i^{(\ell)}),
i=1,\dots,n_\ell,$
be independent copies of
$\left(X^{(\ell)},C^{(\ell)}\right)$, and assume that the samples from
the two groups are independent. We observe the possibly censored
survival times $\widetilde{T}_i^{(\ell)} = \min\{T_i^{(\ell)},C_i^{(\ell)}\}$
and the states
$X_i^{(\ell)}(\widetilde{T}_i^{(\ell)})$ of the patients at
these time points.
The log-likelihood function for group $\ell =1,2$ is given by
\begin{align*}
    \log \mathcal{L}^{(\ell)}(\alpha^{(\ell)},\psi^{(\ell)})
    ={}& \sum_{i = 1}^{n_\ell} \log(S^{(\ell)}(\Tilde{T}_i^{(\ell)})) + \sum_{i = 1}^{n_\ell} \sum_{j = 1}^{k}
    I\{X_i^{(\ell)}(\Tilde{T}_i^{(\ell)}) = j\}
    \log(\alpha_{0j}^{(\ell)}) \\
    &+ \sum_{i = 1}^{n_\ell} I\{X_i^{(\ell)}(\Tilde{T}_i^{(\ell)})=0\}
    \log\left(\frac{g^{(\ell)}(\Tilde{T}_i^{(\ell)},\psi^{(\ell)})}
    {\mathbb{P}(C_i^{(\ell)} \geq \Tilde{T}_i^{(\ell)})}\right) + \sum_{i = 1}^{n_\ell}
    \log(\mathbb{P}(C_i^{(\ell)} \geq \Tilde{T}_i^{(\ell)})).
\end{align*}
Note that under random right censoring the transition intensity MLEs are still given by \eqref{Eq:MLEIntensities}. However, we can now additionally calculate $\psi^{(1)}$ and $\psi^{(2)}$ from the log-likelihood function that is adapted to the random right censoring. For example, for exponentially distributed censoring times $C^{(1)}$ and $C^{(2)}$ with rate parameters $\psi^{(1)}$, $\psi^{(2)}  {>0}$ the MLEs are given by 
\begin{equation*}
    \hat{\psi}^{(\ell)} = \frac{\sum_{i=1}^{n_\ell} I\{X_i^{(\ell)}(\Tilde{T}_i^{(\ell)}) = 0\}}{\sum_{i=1}^{n_\ell} \Tilde{T}_i^{(\ell)}} ~~~(\ell =1,2) .
\end{equation*} 

\subsection{Similarity Test of Competing Risk Models Based on Transition Probabilities} \label{SectionSimilarityTest}

We investigate the similarity of two competing risk models with transition probability functions $\{ P^{(1)}_{0j} \}_{j=1, \ldots ,k}$ and $\{ P^{(2)}_{0j} \}_{j=1, \ldots ,k}$, based on two respective samples, $X_1^{(1)}, \ldots , X_{n_1}^{(1)}$  and $X_1^{(2)}, \ldots , X_{n_2}^{(2)}$ over a time interval $[0,\tau]$ with fixed $\tau \in \mathbb{R}$. In the case of administrative censoring, the natural choice is $\tau = T$, where $T$ is the censoring time. For random right censoring, $\tau$ should be chosen with respect to the research question. Due to the good interpretability of the maximum deviation, we define the similarity measure
\begin{equation} \label{det4}
  d_\infty := \max_{j=1}^k \max_{t \in [0,\tau]}  \big  \vert P^{(1)}_{0j}(t) - P^{(2)}_{0j}(t) \big \vert 
\end{equation}
and propose to test the hypotheses 
\begin{equation} \label{Eq:Hypotheses}
    H_0 : d_\infty \geq \varepsilon_\infty
	\text{ ~~vs. ~~}
    H_1 : d_\infty < \varepsilon_\infty,
\end{equation}
where $\varepsilon_\infty > 0$ is a pre-specified threshold defining when the models are considered as similar (for a more detailed discussion about the choice of $\varepsilon_\infty$ see Remark \ref{rem1}(b)). Note that the alternative $H_1: d_\infty < \varepsilon_\infty$ means that the transition probabilities of model $1$ and $2$ do not differ more than $\varepsilon_\infty$ across all states and at any time before $\tau$.


For these hypotheses, we define in Algorithm \ref{alg1} a test that applies to both types of censoring discussed in Section \ref{Section2.1} and \ref{SectionRandomRightCensoring}, respectively.
\begin{breakablealgorithm} 
    \caption{Similarity Test via Constrained Parametric Bootstrap} \label{alg1}
    \begin{description} 
		\item[$1.$] Calculate the MLEs $\hat{\alpha}^{(1)}, \hat{\alpha}^{(2)}$ defined in \eqref{Eq:MLEIntensities},  in the case of random right censoring, the MLEs of the censoring parameters $\hat{\psi}^{(1)}, \hat{\psi}^{(2)}$, the corresponding estimators of the transition probabilities 
		$\hat{P}_{0j}^{(\ell)} = \big\{ \hat{P}^{(\ell)}_{0j}(t) \vert t \in [0,\tau] \big\}$ for $j=1,\dots,k$ and the test statistic
        \begin{equation*}
            \hat{d}_\infty = \max_{j=1}^k \max_{t \in [0,\tau]} \vert \hat{P}^{(1)}_{0j}(t) - \hat{P}^{(2)}_{0j}(t) \vert.
        \end{equation*}
		\item[$2.$] Calculate the MLEs $\bar{\alpha}^{(1)}$ and $\bar{\alpha}^{(2)}$  of the intensities $\alpha^{(1)}$ and $\alpha^{(2)}$ under the constraint 
		\begin{equation*}
			\max_{j=1}^k \max_{t \in [0,\tau]} \left\vert P^{(1)}_{0j}(t) - P^{(2)}_{0j}(t) \right\vert = \varepsilon_\infty.
		\end{equation*}
      and define constrained estimators $\hat{\hat{\alpha}}^{(\ell)} $
      as  $\hat{\alpha}^{(\ell)}$ if  $\hat{d}_\infty \geq \varepsilon_\infty$ and as $\bar{\alpha}^{(\ell)}$ {if } $\hat{d}_\infty < \varepsilon_\infty$.
      
		\item[$3.1.$] Generate bootstrap data $
			X^{*(1)}_{1}, \dots, X^{*(1)}_{n_1}$ and $X^{*(2)}_{1}, \dots, X^{*(2)}_{n_2}$
		 with parameters  $\hat{\hat{\alpha}}^{(1)}$ and $\hat{\hat{\alpha}}^{(2)}$, respectively. The healthcare pathway of individual $i=1,\dots,n_{\ell}$ from group $\ell=1,2$ is generated by simulating the survival time $T^{*(\ell)}_{i}$ based on $\hat{\hat{\alpha}}^{(\ell)}_0 = \sum_{j=1}^k \hat{\hat{\alpha}}^{(\ell)}_{0j}$ and, by deciding with a multinomial experiment with probabilities $\hat{\hat{\alpha}}^{(\ell)}_{0j} / \hat{\hat{\alpha}}^{(\ell)}_0$ what state the individual transitions to. 
        For administrative censoring, the observed survival time is $\tilde{T}_{i}^{*(\ell)} = \min\{T^{*(\ell)}_{i},T\}$. For random right censoring, generate censoring times $C^{*(\ell)}_i$ from the censoring distribution $G^{(\ell)}(t, \hat{\psi}^{(\ell)})$ and define the observed survival time by $\tilde{T}_{i}^{*(\ell)} = \min\{T^{*(\ell)}_{i},C^{*(\ell)}_i\}$.
        
		\item[$3.2.$] Calculate the MLEs $\hat{\alpha}^{*(1)}$ and $\hat{\alpha}^{*(2)}$ based on the bootstrap data generated in Step $3.1.$ Calculate the corresponding estimators of the transition probabilities  $\hat{P}_{0j} ^{*(\ell)}$ by \eqref{Eq:ProbabilityFromIntensity} 
        with ${\alpha}^{(\ell)}  = \hat{\alpha}^{*(\ell)}$
		and the bootstrap statistic
		\begin{equation} \label{Eq:BootstrapTestStatistic}
			\hat{d}^*_\infty = \max_{j =1}^k \max_{t \in [0,\tau]} \left\vert \hat{P}_{0j}^{*(1)}(t) - \hat{P}_{0j}^{*(2)}(t) \right\vert.
		\end{equation}
		\item[$3.3.$] Repeat the Steps $3.1.$ and $3.2.$ $B$ times to generate $B$ replicates of the test statistic, and denote by $\hat{d}_{\infty(1)}^{*} \leq \dots \leq \hat{d}_{\infty(B)}^{*}$ the order statistic of these replicates.
		\item[$4.$] Calculate the empirical $\alpha$-quantile
		\begin{equation}
			\hat{q}_{\alpha}^{*} := \hat{d}_{\infty(\lfloor \alpha B \rfloor)}^{*}.
		\label{hol1}
        \end{equation}
		\item[$5.$] Reject the null hypothesis in \eqref{Eq:Hypotheses} if
		$
			\hat{d}_\infty < \hat{q}_{\alpha}^{*}.
		$ 
    \end{description}
\end{breakablealgorithm}

\begin{remark} \label{rem1} ~~~

(a) In Theorem~\ref{Th:BootstrapLInfty} of the Supplementary Material, we show that Algorithm \ref{alg1} defines a valid test procedure for the hypotheses in \eqref{Eq:Hypotheses} under administrative censoring as well as random right censoring with exponentially distributed censoring times. Under the null hypothesis the test based on the transition probabilities has asymptotic level $\alpha$. A more precise statement can be made if the set 
	\begin{equation} \label{Def:Epsilon}
				\mathcal{E} = \Big \{ (u,i) \in  [0,\tau] \times \{1,\dots,k\}
                \: \big\vert \: \big \vert P^{(1)}_{0i} (u) -P^{(2)}_{0i} (u) \big \vert
                = d_\infty \Big \}
		\end{equation}
 of all pairs of states and time points where the (absolute) difference between transition probabilities is maximal  has only one element. In this case, 
 the type I error converges to $0$ inside the null hypothesis ($d_\infty > \varepsilon_\infty$) and to the nominal level $\alpha$ on the margin ($d_\infty = \varepsilon_\infty$) as sample sizes $n_1,n_2 \to \infty$ increase. Lastly, the test based on the transition probabilities 
 detects the  alternative ($d_\infty <  \varepsilon_\infty$) with a probability converging to $1$ with increasing sample size, which means that it is consistent. 

The constrained optimization in Step $2$ of Algorithm \ref{alg1} is essential for this result to hold. It guarantees that the bootstrap distribution is generated under the null hypothesis, and in particular under the least favorable null $d_\infty=\varepsilon_\infty$ whenever the unrestricted estimator lies in the alternative. Without this step, the unrestricted bootstrap would still be asymptotically conservative inside the null, but it would not provide the correct margin calibration and would generally fail to yield a consistent similarity test.

(b)
An essential ingredient in testing the hypotheses \eqref{Eq:Hypotheses} is the choice of the threshold $\varepsilon_\infty$, which has to be carefully discussed for each application. 
In situations where it is difficult to fix the threshold in advance we can also define a threshold in a data-driven way. To be precise, note  that the 
  statistic $\hat d^*_\infty$ in \eqref{Eq:BootstrapTestStatistic} depends on the threshold $\varepsilon_\infty$ and in this remark we make this dependence explicit using the notation $\hat d^*_{\infty, \varepsilon_\infty}$.
  Note also that the hypotheses  in \eqref{Eq:Hypotheses} are nested. Recalling the definition of the bootstrap in Algorithm \ref{alg1} it is easy to see that the monotonicity $\hat{d}_{\infty , \varepsilon_1}^{*} \leq \hat{d}_{\infty , \varepsilon_2}^{*}$ whenever $\varepsilon_1 \leq \varepsilon_2$ holds with high probability  if the sample sizes $n_1$ and $n_2$ are sufficiently large. Consequently, we obtain for the corresponding quantiles in \eqref{hol1}
  the inequality $\hat{q}_{\alpha , \varepsilon_1}^{*} \leq \hat{q}_{\alpha , \varepsilon_2}^{*}$ with high probability, and 
rejecting the null hypothesis in \eqref{Eq:Hypotheses} by the test 
for $\varepsilon_\infty = \varepsilon_0$ also yields rejection of the null 
for all $\varepsilon_\infty >\varepsilon_0$. Therefore, by the sequential rejection principle, we may simultaneously test the hypotheses in \eqref{Eq:Hypotheses} for different $\varepsilon_\infty > 0$ starting at a $\varepsilon_\infty$ close to $0$ and increasing $\varepsilon_\infty $ to find the minimum value $ \hat \varepsilon_{\alpha}$ for which $H_0$ is rejected for the first time. This value could be interpreted as a measure of evidence  for similarity with a controlled type I error rate $\alpha$.

\end{remark}

\section{Simulation study} \label{Chapter Simulation}
In this section, we investigate the finite-sample properties of the test proposed in Algorithm \ref{alg1} and compare it with a similarity test recently proposed by \citet{Moellenhoff}, which is based on a similarity measure involving the transition intensities, that is
$$ 
d_{int} := \max_{j=1,\dots,k} \big \vert \alpha_{0j}^{(1)} - \alpha_{0j}^{(2)} \big \vert.
$$
The corresponding hypotheses are therefore defined by 
\begin{equation}
\label{Eq:Hypotheses1}
    H_0: d_{int} \geq \varepsilon_{int} \text{ vs. } H_1: d_{int} < \varepsilon_{int}
\end{equation}
with a threshold $\varepsilon_{int} > 0$. Note that this threshold is not necessarily the same as the threshold $\varepsilon_\infty$ used in the hypotheses \eqref{Eq:Hypotheses} as the hypotheses in \eqref{Eq:Hypotheses1} are based on a different similarity measure. As the similarity measure in \eqref{det4} utilizes the transition probabilities to quantify similarity of competing risks models, we refer to this method as the test based on transition probabilities. The similarity measure $d_{int}$ uses transition intensities to quantify similarity. Thus, we call this method the test based on transition intensities.

Since the two procedures are based on different distances, their rejection probabilities do not constitute a direct power comparison for a common hypothesis. Instead, the simulations compare the finite-sample operating characteristics of two similarity measures under the same data-generating mechanisms. To facilitate the comparison, the thresholds $\varepsilon_{\mathrm{int}}$ and $\varepsilon_\infty$ are calibrated using a common boundary scenario. That means, they are chosen such that the same reference configuration lies on the boundary of the corresponding hypotheses for both procedures. The thresholds are then kept fixed across the remaining scenarios, while the data-generating transition intensities are varied.
This allows us to compare the finite-sample calibration at the common boundary scenario and the empirical rejection behavior of the two procedures across a common set of competing risks models.

We evaluate the performance for administrative censoring and exponential random right censoring. 
In the first setting, individuals who do not transition to any of the competing risks states during a fixed 90-day follow-up period are censored. Thus, we choose the time interval with end point $\tau = 90$ to specify the similarity measure defined in \eqref{det4}. In the second setting, the censoring times $C^{(1)}$, $C^{(2)}$ are exponentially distributed with rate parameters $\psi^{(1)}, \psi^{(2)} > 0$, respectively. To study the effect of different censoring rates, we consider the parameters $\psi^{(1)} = \psi^{(2)} \in \{0.002,0.005,0.01\}$ in the censoring distribution. We opt for $\tau = 90$ in this setting as well.

As the transition intensities uniquely determine a competing risks model and are used to generate data, we define our simulation scenarios in terms of the transition intensities.
As a common boundary scenario, we consider the Margin scenario in Table \ref{tab:transition_intensities}. For these intensities the similarity measure $d_{int}$ is given by $0.0015$, and we choose the threshold $\varepsilon_{int}$ such that the Margin scenario corresponds to the margin for the hypotheses in \eqref{Eq:Hypotheses1}, that is $\varepsilon_{int} = d_{int} = 0.0015$.
To calibrate the test based on transition probabilities, we use \eqref{Eq:ProbabilityFromIntensity} and \eqref{det4} to calculate $d_\infty $ from the Margin scenario in Table \ref{tab:transition_intensities} and define the threshold for the hypotheses \eqref{Eq:Hypotheses} by 
$\varepsilon_\infty = d_{\infty} = 0.1156115779683729$.
Note that this value is rounded to the 16th decimal place to provide high accuracy for the simulations.

To investigate the performance of the tests under administrative censoring, we generate data by simulating healthcare pathways from the Margin transition intensities in Table \ref{tab:transition_intensities} with the algorithm described by \citet{Beyersmann}. The observed survival times are given by the minimum of the generated survival time and $T = 90$. To simulate data inside the null hypothesis, we generate healthcare pathways from transition intensities with a similarity measure $d_\infty$ being larger than the chosen thresholds. To simulate data under the alternative hypothesis, we choose transition intensities with a similarity measure $d_\infty$ being smaller than the chosen thresholds. Overall, we design one scenario inside the null and five scenarios under the alternative, being progressively further away from the margin. All scenarios are described in Table \ref{tab:transition_intensities}, together with the values of the similarity measures $d_\infty$ and $d_{int}$ of the scenarios and their relative similarities with respect to the margin. The ratios $d_\infty/\varepsilon_\infty$ and $d_{\mathrm{int}}/\varepsilon_{\mathrm{int}}$ are not identical for all scenarios. However, they are close enough to compare the empirical rejection behavior of the two tests for scenarios that are approximately equally far inside the respective alternative hypotheses.
\begin{table}
\centering
\small
\caption{Transition intensities $\alpha^{(1)}$ and $\alpha^{(2)}$ to simulate healthcare pathways inside the null hypothesis, on the margin and under the alternative hypothesis.}
\label{tab:transition_intensities}
\resizebox{\textwidth}{!}{%
\begin{tabular}{l|ccc|ccc|c|c|c|c}
\toprule
\rowcolor{gray!20}
\textbf{Scenario} & $\alpha_{01}^{(1)}$ & $\alpha_{02}^{(1)}$ & $\alpha_{03}^{(1)}$ & $\alpha_{01}^{(2)}$ & $\alpha_{02}^{(2)}$ & $\alpha_{03}^{(2)}$ & $d_\infty$ & $d_\infty/ \varepsilon_\infty$ & $d_{int}$ & $d_{int}/\varepsilon_{int}$\\
\midrule
Null          & 0.0025 & 0.0010 & 0.0003 & 0.0007 & 0.0022 & 0.0015 & 0.1385 & 1.1983 & 0.0018 & 1.2000 \\
Margin        & 0.0023 & 0.0011 & 0.0004 & 0.0008 & 0.0021 & 0.0014 & 0.1156 & 1.0000 & 0.0015 & 1.0000 \\
Alternative 1 & 0.0022 & 0.0012 & 0.0005 & 0.0009 & 0.0021 & 0.0014 & 0.1001 & 0.8658 & 0.0013 & 0.8667 \\
Alternative 2 & 0.0021 & 0.0012 & 0.0005 & 0.0010 & 0.0020 & 0.0013 & 0.0854 & 0.7390 & 0.0011 & 0.7333 \\
Alternative 3 & 0.0020 & 0.0013 & 0.0006 & 0.0011 & 0.0019 & 0.0012 & 0.0694 & 0.6000 & 0.0009 & 0.6000 \\
Alternative 4 & 0.0019 & 0.0014 & 0.0007 & 0.0013 & 0.0018 & 0.0011 & 0.0462 & 0.3994 & 0.0006 & 0.4000 \\
Alternative 5 & 0.0017 & 0.0015 & 0.0008 & 0.0014 & 0.0017 & 0.0010 & 0.0231 & 0.2000 & 0.0003 & 0.2000 \\
\bottomrule
\end{tabular}
}
\end{table}

To evaluate the performance of the similarity test based on transition probabilities for random right censoring, we choose the same scenarios as described in Table \ref{tab:transition_intensities}. In addition to the event times, we generate censoring times from the Exp$(\psi^{(\ell)})$-distribution. The observed survival time is the minimum of the survival time and censoring time.

We investigate the performance of our method and compare it to the method of \citet{Moellenhoff} for a wide variety of sample sizes, i.e. $(n_1,n_2) \in \{(50,50)$, $(100,100)$, $(200,200)$, $(300,300)$,$(250,450)$, $(500,500)\}$.
The choice of $(n_1,n_2) = (250,450)$ is close to the observed sample size in the real data example in Section \ref{Section 4} and enables us to analyze the effect of unbalanced sample sizes. Furthermore, the first four sample sizes $(50,50)$, $(100,100)$, $(200,200)$, $(300,300)$ correspond to settings where we observe even fewer competing risks events than in the real data set.

The percentage of censored individuals changes depending on the scenario from Table \ref{tab:transition_intensities}. The scenarios are constructed in a way that keeps the all-cause hazard relatively stable across the data-generating mechanisms resulting in comparatively stable proportions of censored individuals. For Exp$(0.002)$ censoring, the percentage of censored individuals lies between $31.3 \%$ and $34.5 \%$ corresponding to light to medium censoring. The Exp$(0.005)$ censoring is medium to heavy with $53.2 \%$ to $56.8 \%$ of censored patients. Exp$(0.01)$ and administrative censoring can be assigned to heavy censoring with $69.4 \%$ to $72.5 \%$ and $67.3 \%$ to $71.0 \%$ of censored individuals, respectively.
The finite-sample performance of the tests is driven by the amount of observed information reflected by the censoring proportion as well as by the number of transitions and the total time at risk. To visualize the limited amount of available information the expected number of transitions for group $1$ and group $2$ is listed in Table~\ref{Tab:ExpectedNumberOfTransitionsStates} of the Supplementary Material for some settings. For example, in the null scenario for $(n_1,n_2)=(250,450)$, the expected total number of transitions into state $3$ for group $1$ is $5.4$ under Exp$(0.01)$ censoring while the total number of expected transitions in this group is $68.8$.

The performance of the test based on transition probabilities is assessed in terms of type I error and power that are documented in Table \ref{prop_rejection_margin_null} of this paper and Table~\ref{prop_rejection_alt_with_brackets} of the Supplementary Material. The first entry in each table represents the empirical rejection probabilities of the testing procedure presented in this article (Algorithm \ref{alg1}). The number in brackets corresponds to the empirical rejection probabilities of the test suggested by \citet{Moellenhoff}. All tests are performed with a nominal level of $\alpha = 0.05$ and unless stated otherwise, $B = 1000$ bootstrap repetitions and $2000$ simulation runs are used for each scenario to calculate the empirical rejection probabilities.

\subsection{Type I error}
Table~\ref{prop_rejection_margin_null} summarizes the simulated type I error. First, note that the set $\mathcal{E}$ defined in \eqref{Def:Epsilon} has cardinality $1$ in all cases under consideration and we expect (by the discussion in Remark \ref{rem1}(a)) that for an increasing sample size the empirical rejection probability inside the null hypothesis should converge to $0$ while it converges to $0.05$ on the margin. For the test based on transition intensities, we expect the same behavior if $\max_{j=1,\dots,k} |\alpha_{0j}^{(1)}-\alpha_{0j}^{(2)}|$ is obtained in exactly one state $j$. The transition intensities in Table \ref{tab:transition_intensities} are chosen such that this is the case for all scenarios.

The empirical type \uproman{1} error of the test based on transition probabilities inside the null hypothesis varies between $0\%$ and $0.7\%$. At a relative similarity distance of approximately $1.2$, the test rarely rejects the null hypothesis of dissimilarity across all sample sizes and censoring mechanisms. Since we observe similar rejection rates for the test based on transition intensities it can be said that both tests are strongly conservative inside the null hypothesis.

At the boundary, the test based on transition probabilities is conservative for small sample sizes with rejection rates between $2\%$ and $3.95\%$ and approaches the nominal level as the sample size increases. For instance, the type I error on the margin for Exp$(0.002)$ censoring increases from $0.02$ for a sample size of $(50,50)$ to $0.0575$ for a sample size of $(500,500)$. For the largest sample sizes, the empirical rejection probabilities are close to 0.05, with occasional mild exceedances, such as $0.064$ for $(n_1,n_2)=(500,500)$ under Exp$(0.005)$ censoring.
Similar behavior at the boundary is observed for the benchmark test based on transition intensities where a mild exceedance can be observed for $(n_1,n_2)=(250,450)$ under Exp$(0.002)$ censoring.

Stronger censoring reduces the effective amount of information through fewer observed transitions and less total time at risk. This slows down convergence to the asymptotic boundary distribution and makes the constrained bootstrap test conservative in finite samples. Consequently, we observe that the empirical rejection probability at the margin is smaller under heavier censoring compared to light censoring. For $(n_1,n_2) = (200,200)$ the rejection rate is 0.0260 in the case of administrative censoring while it is $0.0525$ for Exp$(0.002)$ censoring. This effect vanishes as the sample size, and hence the amount of observed information, increases.

\begin{table}
    \caption{Empirical rejection probabilities of the similarity test based on transition probabilities under administrative and exponential random right censoring on the margin (Margin) and inside the null hypothesis (Null) for different censoring mechanisms and different sample sizes. The numbers in brackets correspond to the empirical rejection probabilities of the similarity test based on transition intensities \citep{Moellenhoff}. The nominal level is $\alpha = 0.05$.}
    \label{prop_rejection_margin_null}
    \centering
    \small
    \begin{tabular}{c c c c}
        \toprule
        \shortstack{Sample size} & \shortstack{Censoring} 
        & Margin & Null \\
        \midrule

        \multirow{4}{*}{(50, 50)}  
            & Exp(0.002) & 0.0200 (0.0185) & 0.0040 (0.0025) \\
            & Exp(0.005) & 0.0200 (0.0170) & 0.0050 (0.0060) \\
            & Exp(0.01)  & 0.0180 (0.0195) & 0.0070 (0.0070) \\
            & adm       & 0.0230 (0.0215) & 0.0065 (0.0065) \\
        \hline
        \multirow{4}{*}{(100, 100)}  
            & Exp(0.002) & 0.0395 (0.0290) & 0.0035 (0.0020) \\
            & Exp(0.005) & 0.0290 (0.0205) & 0.0040 (0.0035) \\
            & Exp(0.01)  & 0.0210 (0.0175) & 0.0060 (0.0060) \\
            & adm       & 0.0275 (0.0215) & 0.0055 (0.0065) \\
        \hline
        \multirow{4}{*}{(200, 200)}  
            & Exp(0.002) & 0.0525 (0.0470) & 0.0035 (0.0020) \\
            & Exp(0.005) & 0.0415 (0.0390) & 0.0020 (0.0025) \\
            & Exp(0.01)  & 0.0385 (0.0255) & 0.0045 (0.0030) \\
            & adm       & 0.0260 (0.0245) & 0.0060 (0.0025) \\
        \hline
        \multirow{4}{*}{(300, 300)}  
            & Exp(0.002) & 0.0585 (0.0520) & 0.0015 (0.0020) \\
            & Exp(0.005) & 0.0525 (0.0450) & 0.0020 (0.0020) \\
            & Exp(0.01)  & 0.0490 (0.0365) & 0.0035 (0.0045) \\
            & adm       & 0.0515 (0.0410) & 0.0070 (0.0035) \\
        \hline
        \multirow{4}{*}{(250, 450)}  
            & Exp(0.002) & 0.0610 (0.0660) & 0.0030 (0.0035) \\
            & Exp(0.005) & 0.0610 (0.0530) & 0.0055 (0.0075) \\
            & Exp(0.01)  & 0.0540 (0.0465) & 0.0055 (0.0055) \\
            & adm       & 0.0545 (0.0530) & 0.0065 (0.0065) \\
        \hline
        \multirow{4}{*}{(500, 500)}  
            & Exp(0.002) & 0.0575 (0.0605) & 0.0005 (0.0010) \\
            & Exp(0.005) & 0.0640 (0.0530) & 0.0020 (0.0030) \\
            & Exp(0.01)  & 0.0550 (0.0470) & 0.0030 (0.0030) \\
            & adm       & 0.0570 (0.0520) & 0.0025 (0.0020) \\
        \bottomrule
    \end{tabular}
\end{table}

Additional simulations with group sizes $1000$ and $2000$, reported in Table~\ref{prop_rejection_margin_high} of the Supplementary Material, further support convergence of both procedures to the nominal level at the boundary. For example, for $(n_1,n_2) = (2000,2000)$ the type \uproman{1} error under Exp$(0.005)$ censoring is $0.0490$ for the test based on transition probabilities and $0.0520$ for the test based on transition intensities.

\subsection{Power}
\begin{figure}
    \centering
    \includegraphics[width=0.68\textwidth]{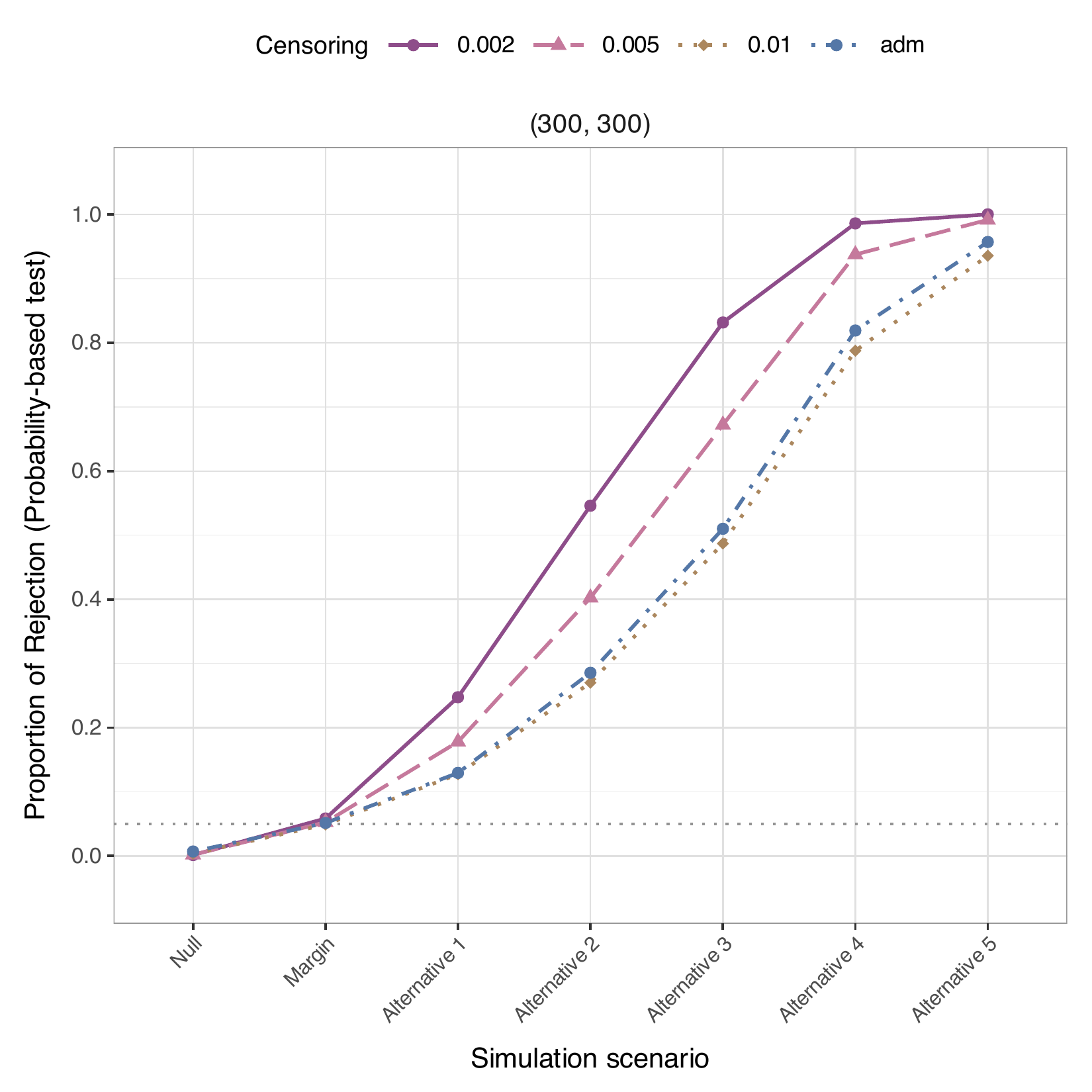}
    \caption{Empirical rejection probabilities of the similarity test based on transition probabilities for the scenarios in Table \ref{tab:transition_intensities} for $(n_1,n_2) =(300,300)$. The different lines represent the rejection rate for different censoring mechanisms.}
    \label{Power_vs_Alternative}
\end{figure}

Figure \ref{Power_vs_Alternative} shows the empirical rejection probabilities of the test based on transition probabilities for $(n_1,n_2)=(300,300)$ across all considered censoring mechanisms. It shows that a higher proportion of censored individuals leads to a loss of power. The effect of censoring is most pronounced for the intermediate Alternatives 2 and 3. For example, for alternative $3$, the power drops from $0.8315$ at exponential censoring rate $0.002$ to $0.4870$ at rate $0.01$. The proportion of administratively censored individuals lies always just below the censored proportion for Exp$(0.01)$ censoring while the power in the administratively censored case is slightly larger than in the Exp$(0.01)$ censored case. For alternative $3$, this means a rejection rate of $0.5100$. The similar power curves under administrative and Exp$(0.01)$ censoring suggest that for these censoring mechanisms the finite-sample performance is primarily driven by the resulting total time at risk rather than by the particular form of the censoring mechanism.

For every censoring mechanism, the rejection
probability increases as the alternatives move farther away from the boundary of the null hypothesis. We observe that it is difficult to distinguish Alternative $1$ from the boundary across all censoring mechanisms, while Alternative $2$ still yields only moderate power.
Alternative $3$, for which $d_\infty / \varepsilon_\infty = 0.6$, represents an intermediate scenario, where power increases clearly with the sample size, but continues to depend substantially on the amount of censoring. Alternative $3$ therefore marks the transition between alternatives close to the margin and the stronger Alternatives $4$ and $5$, for which high power is obtained more uniformly across censoring mechanisms. Under Alternative 5, power exceeds $0.93$ for every censoring mechanism, even though the power for mechanisms yielding smaller proportions of censored individuals tends to $1$ faster.
 The complete power results for all sample sizes and alternatives provided in Table~\ref{prop_rejection_alt_with_brackets} of the Supplementary Material show similar trends across all sample sizes. In Section~\ref{supp:additional_results} of the Supplementary Material we discuss the results above with greater focus on the influence of the samples size.
Additional power curves as a function of sample size are shown in Figure~\ref{Power_vs_SampleSize_prob_Alt3,4} of the Supplementary Material for alternative $3$ and $4$. These curves especially visualize that the test based on transition probabilities remains stable under moderately unbalanced group sizes.

\begin{figure}
    \centering
    \includegraphics[width=0.78\textwidth]{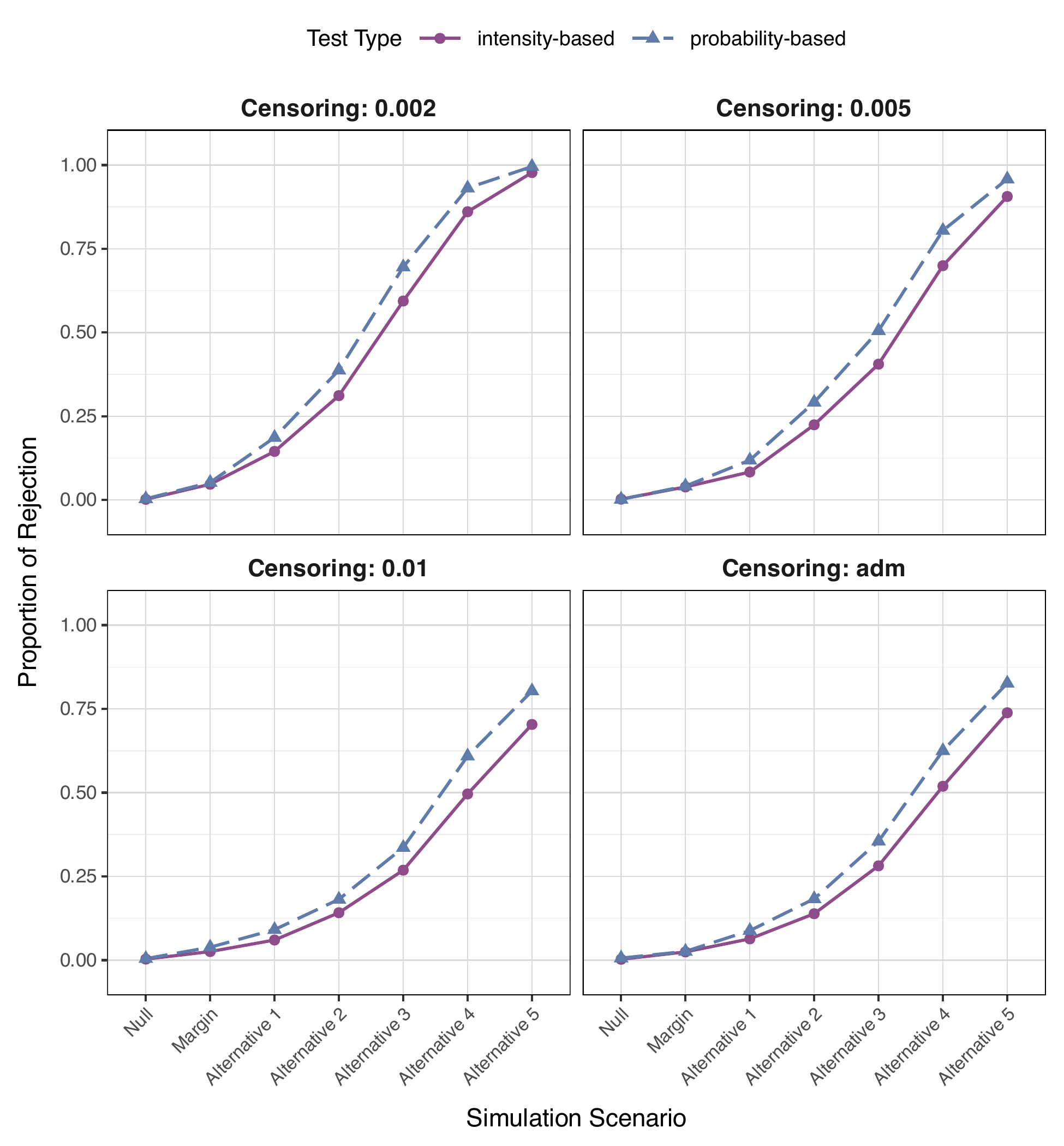}
    \caption{
    Empirical rejection probabilities of the similarity test based on transition probabilities and the test based on transition intensities under the scenarios from Table \ref{tab:transition_intensities}. The sample size is $(n_1,n_2) = (200,200)$ and four censoring scenarios are considered.}
    \label{Intensity_vs_Prob}
\end{figure}

We conclude this section with a comparison of the test based on transition probabilities and the test based on transition intensities.
Figure \ref{Intensity_vs_Prob} compares the tests for all scenarios from Table \ref{tab:transition_intensities} and all censoring mechanisms for sample size $(n_1,n_2)=(200,200)$. The test based on transition probabilities has larger
empirical rejection probabilities under all displayed alternatives. For example, for Exp$(0.002)$ censoring, the gain in rejection probability under Alternative $3$ is $0.1015$, increasing from $0.5940$ for the test based on transition intensities to $0.6955$ for the test based on transition probabilities. Note that Alternative $3$ represents an equally extreme alternative on the transition intensity scale and the transition probability scale because it corresponds to a relative distance with respect to the threshold of $0.6$ on both scales. The difference between the two procedures is most pronounced for intermediate power values, where the rejection probabilities are above $0.5$ but not yet close to $1$. Under the strong Alternative $5$, both tests attain nearly full power for Exp$(0.002)$ censoring, although the test based on transition probabilities reaches large rejection probabilities earlier. Table~\ref{prop_rejection_alt_with_brackets} of the Supplementary Material shows that the empirical rejection probabilities of the test based on transition probabilities exceed those of the benchmark test based on transition intensities in nearly all other combinations of scenario, censoring mechanism and sample size as well.
These results suggest that, for the considered data-generating mechanisms, defining similarity through transition probabilities instead of transition intensities can lead to a noticeable gain in finite-sample rejection probability across comparable alternatives.

\section{Application: Healthcare Pathways of Prostate Cancer Patients} \label{Section 4}
In our application example, we analyze routine inpatient data of prostate cancer patients treated at the Department of Urology, University Medical Center Freiburg, between 1 January 2015 and 1 February 2021. All patients underwent open radical prostatectomy (ORP), involving removal of the prostate and seminal vesicles. The dataset comprises two groups: those who received in-house MRI-based fusion biopsy (FB) before surgery and those who did not.
We aim to assess whether post-surgical pathways, specifically regarding probabilities of short-term hospital readmission, are similar between these groups to evaluate whether data can be pooled. In total, 213 patients (31\%) had in-house FB, while 482 (69\%) did not.
Prostate cancer is among the most common reasons for inpatient admission, and ORP is a standard treatment. While FB is the current diagnostic standard, not all patients undergo it before surgery, depending on the referring urologist’s practice.
Hospital readmissions were categorized into three groups according to their main ICD-10 diagnosis, ``I89.8: Other specified noninfective disorders of lymphatic vessels and lymph nodes'', ``C61: Malignant neoplasm of prostate'', and a combined category for ``any other diagnosis''. As the most common postoperative complications of ORP develop within a few weeks after surgery, we consider administrative censoring after 90 days in our analysis. Within this period, readmissions occurred for ICD-10 I89.8 (Model 1: n = 17 [8\%], Model 2: n = 29 [6\%]), ICD-10 C61 (Model 1: n = 18 [8\%], Model 2: n = 60 [12\%]), and ``any other diagnosis'' (Model 1: n = 6 [3\%], Model 2: n = 31 [6\%]). 
Readmissions occurring within the 90-day period were modeled using a competing risks modeling framework, with the initial states corresponding to ``ORP with prior in-house FB'' and ``ORP without prior in-house FB'' for the two groups, respectively. The absorbing states are defined by the three readmission categories.

For this application, transition probabilities correspond directly to the clinically relevant quantities. That is $P_{0j}^{(\ell)}(t)$ is the probability that a readmission of type $j$ has occurred by time $t$ in group $\ell$. By contrast, the corresponding transition intensity describes the instantaneous rate of this readmission among patients who have not yet experienced any of the modeled events.
Using the plug-in formula for the transition probability estimators described in \eqref{Eq:ProbabilityFromIntensity}, we first examine the estimated transition probability functions $\hat{P}_{0j}^{(1)}$ and $\hat{P}_{0j}^{(2)}$.
Across all event types $j$, the absolute differences $|\hat{P}_{0j}^{(1)}(t) - \hat{P}_{0j}^{(2)}(t)|$ increase  over the whole 90-day follow-up period. The largest discrepancy is observed for readmissions due to ICD-10 C61, where the absolute difference reaches 0.0432 at 90 days. Differences for ICD-10 I89.8 and for other diagnoses are overall smaller with $0.0186$ and $0.0375$, respectively. These observed magnitudes provide a descriptive reference for interpreting the similarity thresholds considered in the formal testing procedure below.


To formally assess similarity, we test the hypotheses in \eqref{Eq:Hypotheses}. Rejecting the null hypothesis provides evidence that none of the three cause-specific readmission probabilities differs between the groups by $\varepsilon_\infty$ or more at any time during the $90$-day follow-up period.

Because no established clinical similarity margin is available for this application, Table \ref{pvalue} reports the results across a range of margins (see Remark \ref{rem1}(b)). Among the considered values, $0.087$ is the smallest margin for which the null hypothesis is rejected at the level $0.05$. Thus, under the fitted model, the data provide evidence that the maximal difference between the two groups’ readmission probabilities is smaller than $8.7 \%$. By contrast, similarity cannot be established at margins of $0.08$ or smaller on the basis of these data.

Whether a difference of $8.7 \%$ is sufficiently small remains a clinical judgment rather than a conclusion determined by the statistical procedure. If this margin is considered clinically acceptable, the results may support combining the groups for subsequent analyses specifically concerned with these $90$-day readmission dynamics.

\begin{table}
\caption{P-values of the similarity test based on transition probabilities described in Algorithm \ref{alg1} for the application example above considering different thresholds $\varepsilon_\infty$. The bold numbers mark p-values below the nominal level of $\alpha = 0.05$.}
\label{pvalue}
\centering
\renewcommand{\arraystretch}{1.3}
\resizebox{\textwidth}{!}{%
\begin{tabular}{lllllllllll}
\toprule
Threshold $\varepsilon_\infty$ 
& 0.05 & 0.06 & 0.07 & 0.08 & 0.087 & 0.09 & 0.10 & 0.11 & 0.12 & 0.13 \\
\midrule
P-value 
& 0.362 & 0.268 & 0.160 & 0.090 & \textbf{0.044} &\textbf{0.034} & \textbf{0.006} & \textbf{0.002} & \textbf{0.002} & \textbf{0.000} \\
\bottomrule
\end{tabular}
}
\end{table}


\section{Discussion}

In this article, we develop a novel parametric bootstrap test to evaluate the similarity of two competing risks models with constant transition intensities. In contrast to previous work on similarity testing, which assesses the similarity of competing risks models in terms of transition intensities \citep{Binder,Moellenhoff}, our approach measures similarity between transition probabilities. This makes the results of the test easier to interpret in a medical context, where outcomes are typically assessed in terms of cumulative risks rather than instantaneous event rates.
For the sake of comparability with the approach of \citet{Moellenhoff}, we opted for a similarity measure based on the maximum distance. The choice of the similarity measure ultimately depends on the research question and exploring alternative distances based on transition probabilities remains an important branch of future research. 

The performance of the proposed similarity test based on transition probabilities was evaluated through extensive simulation studies across a wide range of settings, including varying sample sizes, censoring rates, and degrees of model dissimilarity. The results support the asymptotic level properties that the test is conservative inside the null hypothesis and approaches the nominal level at the margin as the sample size increases. Across the considered alternatives, the test based on transition probabilities yields larger finite-sample rejection probabilities than the transition-intensity benchmark of \citet{Moellenhoff}, especially for intermediate alternatives.

 This increase in empirical rejection probabilities under comparable alternatives comes at the cost of additional complexity in establishing theoretical results for similarity measures based on transition probabilities.  In principle, the proposed procedure can also be formulated for general parametric transition-intensity models.  However, deriving rigorous statistical guarantees directly in this broader setting is substantially more difficult. In particular, establishing the asymptotic distribution of the test statistic requires a detailed analysis of the differentiability properties of the mapping from transition intensities to transition probabilities, which becomes technically involved when the transition intensities are non-constant. Beyond the theoretical challenges, extending the algorithm to general parametric transition-intensity models requires repeatedly solving constrained optimization problems formulated in terms of transition probabilities that typically lack closed-form expressions, thereby making the numerical implementation substantially more demanding. The assumption of constant transition intensities provides a controlled setting in which to assess the methodological differences between intensity- and probability-based similarity measures. This setting allows us to isolate the effect of the chosen similarity measure from the additional complexity introduced by more flexible intensity models.
 Addressing the substantial mathematical challenges associated with non-constant intensities and extending the theoretical guarantees to competing risks models with parametric transition intensities constitute important directions for future research. A further objective is to broaden the framework to more general multistate models with parametrically specified transition intensities.

We may further note here that potential model misspecification is always an important issue when it comes to parametric modeling. While we do not investigate robustness of our approach to model misspecification in this manuscript, \citet{Moellenhoff} studied this issue extensively in dedicated simulations and reported that moderate misspecification had only minor impact, whereas stronger misspecification could lead to either mild type \uproman{1} error inflation or conservative behavior with reduced power. Given that our approach uses the same model class for event dynamics, we expect a qualitatively similar pattern.

Finally, we illustrated the practical utility of our method through a real-world application involving prostate cancer patients who underwent radical prostatectomy with or without prior in-house MRI-targeted fusion biopsy. The analysis focused on comparing the 90-day hospital readmission patterns between the two groups across multiple causes. By applying our test based on transition probabilities, we were able to determine threshold values above which the global null hypothesis of dissimilarity could be rejected. Given the lack of established clinical guidelines for selecting similarity thresholds, further methodological and application-focused research on threshold specification is warranted.

In this paper we were particularly interested in data analysis in small data settings, where nonparametric methods may be less stable. However, an interesting direction of future research is the development of nonparametric methodology for the similarity problems considered in this paper, which can be used without specific model assumptions if a sufficient amount of data is available. 
As the cumulative intensities and transition probabilities can be estimated non-parametrically via the Nelson-Aalen and Aalen-Johansen estimators, respectively, \citep[see, for example,][]{AndersenCountingProcesses}, we expect that corresponding methodology can be developed using similar techniques as described in Section 2.4 of \citet{DetteNeumeyer}.

\bibliographystyle{oup-abbrvnat}
\bibliography{SimilarityTesting}

\section*{Statements and declarations}

\paragraph{Ethical considerations.}
The protocol for the analysis of data in this work was approved by the Ethics Committee, Medical Center – University of Freiburg, Freiburg, Germany (reference number 22-1433-S1-retro). Additional approval for data use was obtained by the Use \& Access Committee (UAC) of the Medical Center – University of Freiburg.

\paragraph{Consent to participate}
The Ethics Committee, Medical Center – University of Freiburg, Freiburg, Germany, waived the requirement for informed consent because the data were analyzed anonymously. 

\paragraph{Funding.}
This work was funded by the Deutsche Forschungsgemeinschaft (DFG, German Research Foundation), Project-ID 499552394, SFB 1597. Holger Dette was supported by the European Union through the European Joint Programme on Rare Diseases under the European Union's Horizon 2020 Research and Innovation Programme, Grant Agreement Number 825575.

\paragraph{Declaration of conflicting interests.}
The authors declared no potential conflicts of interest with respect to the research, authorship and/or publication of this article.

\paragraph{Data and code availability.}
The code used in the simulation study and a tutorial are available at
\url{https://github.com/ZoeLange/Testing-Similarity-of-Competing-Risks-Models-by-Comparing-Transition-Probabilities}.
The clinical routine-care data analyzed in this study are not publicly available because of data-protection and institutional regulations.

\clearpage
\section*{Supplementary Material}
\addcontentsline{toc}{section}{Supplementary Material}
\suppressfloats[t]

\setcounter{section}{0}
\setcounter{subsection}{0}
\setcounter{equation}{0}
\setcounter{figure}{0}
\setcounter{table}{0}
\setcounter{theorem}{0}

\renewcommand{\thetheorem}{S\arabic{theorem}}
\renewcommand{\theequation}{S\arabic{equation}}
\renewcommand{\thefigure}{S\arabic{figure}}
\renewcommand{\thetable}{S\arabic{table}}
\renewcommand{\thesection}{S\arabic{section}}
\renewcommand{\theHsection}{supp.\arabic{section}}
\renewcommand{\theHsubsection}{supp.\arabic{section}.\arabic{subsection}}
\renewcommand{\theHequation}{supp.\arabic{equation}}
\renewcommand{\theHfigure}{supp.\arabic{figure}}
\renewcommand{\theHtable}{supp.\arabic{table}}
\renewcommand{\theHtheorem}{supp.\arabic{theorem}}

\section{Additional Results from the Simulation Study}
\label{supp:additional_results}
\subsection{Expected Number of Transitions}
Table \ref{Tab:ExpectedNumberOfTransitionsStates} illustrates the amount of available information for the estimation of the test statistic by showing the expected number of transitions for the null scenario from Table~\ref{tab:transition_intensities} and for the sample sizes $(100,100)$ and $(250,450)$. For example, for $(n_1,n_2)=(100,100)$ under administrative censoring, the expected total number of transitions in group $1$ is $29.0$, with only $2.3$ expected transitions into state $3$.

Similarly, in the null scenario for $(n_1,n_2)=(250,450)$, the expected total number of transitions in group $1$ decreases from $163.8$ under weak Exp$(0.002)$ censoring to $68.8$ under stronger Exp$(0.01)$ censoring, while $12.9$ to $5.4$ transitions into state $3$ are expected. For group $2$, the corresponding total expected number of transitions decreases from $309.4$ to $137.6$. Thus, some state- and group-specific transition intensities are estimated from very few observed events even when the overall sample size is not particularly small.

\begin{table}
\caption{Expected number of transitions for each state for administrative and random right censoring for the null scenario and sample sizes $(100,100)$ and $(250,450)$.}
\label{Tab:ExpectedNumberOfTransitionsStates}
\centering
\small
\begin{tabular}{c c r r r r r r}
\toprule
\shortstack{Sample \\ size}
& \shortstack{Censoring \\ mechanism}
& \shortstack{Group 1 \\ State 1}
& \shortstack{Group 1 \\ State 2}
& \shortstack{Group 1 \\ State 3}
& \shortstack{Group 2 \\ State 1}
& \shortstack{Group 2 \\ State 2}
& \shortstack{Group 2 \\ State 3} \\
\midrule
    \multirow{4}{*}{(100,100)} 
        & Exp$(0.002)$ 
        & 43.1 & 17.2 & 5.2 & 10.9 & 34.4 & 23.4 \\
        & Exp$(0.005)$ 
        & 28.4 & 11.4 & 3.4 & 7.4 & 23.4 & 16.0 \\
        & Exp$(0.01)$ 
        & 18.1 & 7.2 & 2.2 & 4.9 & 15.3 & 10.4 \\
        & adm
        & 19.1 & 7.6 & 2.3 & 5.2 & 16.3 & 11.1 \\
    \hline

    \multirow{4}{*}{(250,450)} 
        & Exp$(0.002)$ 
        & 107.8 & 43.1 & 12.9 & 49.2 & 154.7 & 105.5 \\
        & Exp$(0.005)$ 
        & 71.0 & 28.4 & 8.5 & 33.5 & 105.3 & 71.8 \\
        & Exp$(0.01)$ 
        & 45.3 & 18.1 & 5.4 & 21.9 & 68.8 & 46.9 \\
        & adm
        & 47.6 & 19.1 & 5.7 & 23.4 & 73.6 & 50.2 \\
    \bottomrule
\end{tabular}
\end{table}

\subsection{Type I error}
The additional simulations in Table \ref{prop_rejection_margin_high} provide further evidence for the asymptotic calibration at the boundary. For $(n_1,n_2)=(1000,1000)$, the
rejection probabilities of the test based on transition probabilities range from
$0.0486$ to $0.0594$ across the censoring mechanisms. For
$(n_1,n_2)=(2000,2000)$, they range from $0.0488$ to $0.0546$. The test based on transition intensities shows a comparable pattern. Thus, the pronounced
conservativeness observed for some smaller sample sizes is no longer apparent at these larger group sizes, and the results are consistent with the convergence to the nominal level stated in Theorem \ref{Th:BootstrapLInfty}.

\begin{table}
    \caption{Empirical rejection probabilities of the similarity test based on transition probabilities under administrative and exponential random right censoring on the margin (Margin) for different censoring mechanisms and high sample sizes. The numbers in brackets correspond to the empirical rejection probabilities of the similarity test based on transition intensities \citep{Moellenhoff}. The nominal level is $\alpha = 0.05$, $B=2000$, $N=5000$.}
    \label{prop_rejection_margin_high}
    \centering
    \small
    \begin{tabular}{c c c}
        \toprule
        \shortstack{Sample size} & \shortstack{Censoring} 
        & Margin \\
        \midrule

        \multirow{4}{*}{(1000, 1000)}  
            & Exp(0.002) & 0.0486 (0.0538) \\
            & Exp(0.005) & 0.0532 (0.0616) \\
            & Exp(0.01)  & 0.0528 (0.0552) \\
            & adm       & 0.0594 (0.0586) \\
        \hline
        \multirow{4}{*}{(2000, 2000)}  
            & Exp(0.002) & 0.0488 (0.0468) \\
            & Exp(0.005) & 0.0490 (0.0520) \\
            & Exp(0.01)  & 0.0546 (0.0592) \\
            & adm       & 0.0490 (0.0528) \\
        \bottomrule
    \end{tabular}
\end{table}

\subsection{Power}
The power of the test based on transition probabilities increases consistently with the group sizes, although the rate of increase depends strongly on the distance from the similarity margin and on the censoring mechanism. For alternatives close to the margin, the increase is
comparatively slow. In particular, under Alternative $1$, the empirical power remains below $0.3195$ even
for $(n_1,n_2)=(500,500)$. By contrast, the gains from increasing the sample size are particularly pronounced under the intermediate Alternatives $2$
and $3$. For the stronger
Alternatives $4$ and $5$, high power is attained at smaller sample sizes, and further increases in the group sizes provide progressively smaller gains as the rejection probabilities approach one. Under Exp$(0.01)$ censoring, the smallest considered balanced group sizes at which the empirical power of the test based on transition probabilities exceeds $0.7$ are $(500,500)$ for alternative $3$, $(300,300)$ for alternative $4$ and $(200,200)$ for Alternative $5$. Thus, this power is reached for a total expected number of transitions for both groups together of $288.2$, $174.4$ and $115.3$ with the smallest expected number per group and state given by $21.6$, $15.0$ and $11.4$, respectively.

The results also show that heavier censoring effectively shifts the power development towards larger sample sizes. This effect is most apparent for intermediate alternatives, whereas it is less visible close to the margin, where power remains low, and under strong alternatives, where power approaches one. The test based on transition intensities exhibits the same general dependence on the sample size, although its rejection probabilities are lower in most of the considered settings.

\begin{table}
    \caption{Empirical rejection probabilities of the similarity test based on transition probabilities for administrative and exponential random right censoring under different alternatives and for different censoring mechanisms and different sample sizes. The numbers in brackets correspond to the empirical rejection probabilities of the test based on transition intensities \citep{Moellenhoff}. The nominal level is $\alpha = 0.05$.}
    \label{prop_rejection_alt_with_brackets}
    \centering
    \small
    \resizebox{\textwidth}{!}{%
    \begin{tabular}{c c c c c c c}
        \toprule
        \shortstack{Sample  \\ size} & 
        \shortstack{Censoring \\ mechanism} & 
        Alt1 & Alt2 & Alt3 & Alt4 & Alt5 \\
        \midrule
        
        \multirow{4}{*}{(50, 50)}  
            & Exp(0.002) & 0.0475 (0.0360) & 0.0980 (0.0775) & 0.1905 (0.1505) & 0.3435 (0.2715) & 0.5000 (0.4020) \\
            & Exp(0.005) & 0.0455 (0.0375) & 0.0770 (0.0660) & 0.1455 (0.1260) & 0.2415 (0.2020) & 0.3225 (0.2650) \\
            & Exp(0.01)  & 0.0380 (0.0385) & 0.0605 (0.0600) & 0.1045 (0.0920) & 0.1400 (0.1260) & 0.1940 (0.1635) \\
            & adm       & 0.0360 (0.0290) & 0.0530 (0.0470) & 0.0955 (0.0775) & 0.1575 (0.1355) & 0.2075 (0.1655) \\
        \hline
        \multirow{4}{*}{(100, 100)}  
            & Exp(0.002) & 0.0990 (0.0705) & 0.2050 (0.1580) & 0.3935 (0.3125) & 0.6615 (0.5450) & 0.8615 (0.7660) \\
            & Exp(0.005) & 0.0640 (0.0495) & 0.1320 (0.1000) & 0.2640 (0.2025) & 0.4675 (0.3730) & 0.6730 (0.5685) \\
            & Exp(0.01)  & 0.0380 (0.0270) & 0.0860 (0.0720) & 0.1580 (0.1235) & 0.3000 (0.2430) & 0.4410 (0.3590) \\
            & adm       & 0.0490 (0.0385) & 0.0880 (0.0725) & 0.1880 (0.1475) & 0.3050 (0.2445) & 0.4915 (0.3980) \\
        \hline
        \multirow{4}{*}{(200, 200)}  
            & Exp(0.002) & 0.1865 (0.1450) & 0.3875 (0.3115) & 0.6955 (0.5940) & 0.9310 (0.8605) & 0.9955 (0.9775) \\
            & Exp(0.005) & 0.1190 (0.0835) & 0.2915 (0.2245) & 0.5055 (0.4055) & 0.8045 (0.6995) & 0.9580 (0.9060) \\
            & Exp(0.01)  & 0.0910 (0.0600) & 0.1815 (0.1415) & 0.3360 (0.2685) & 0.6090 (0.4960) & 0.8035 (0.7035) \\
            & adm       & 0.0880 (0.0635) & 0.1830 (0.1385) & 0.3550 (0.2815) & 0.6250 (0.5190) & 0.8265 (0.7385) \\
        \hline
        \multirow{4}{*}{(300, 300)}  
            & Exp(0.002) & 0.2475 (0.1960) & 0.5460 (0.4570) & 0.8315 (0.7585) & 0.9860 (0.9600) & 1.0000 (0.9990) \\
            & Exp(0.005) & 0.1780 (0.1260) & 0.4025 (0.3390) & 0.6720 (0.5635) & 0.9375 (0.8585) & 0.9915 (0.9700) \\
            & Exp(0.01)  & 0.1275 (0.0950) & 0.2700 (0.2185) & 0.4870 (0.3900) & 0.7875 (0.6790) & 0.9355 (0.8735) \\
            & adm       & 0.1295 (0.0920) & 0.2855 (0.2275) & 0.5100 (0.4070) & 0.8190 (0.7095) & 0.9570 (0.9045) \\
        \hline
        \multirow{4}{*}{(250, 450)}  
            & Exp(0.002) & 0.2555 (0.2355) & 0.5225 (0.4785) & 0.8165 (0.7510) & 0.9800 (0.9510) & 0.9995 (0.9965) \\
            & Exp(0.005) & 0.2030 (0.1790) & 0.4115 (0.3625) & 0.6935 (0.6075) & 0.9260 (0.8615) & 0.9930 (0.9820) \\
            & Exp(0.01)  & 0.1580 (0.1255) & 0.2950 (0.2395) & 0.5090 (0.4270) & 0.7980 (0.7010) & 0.9415 (0.8840) \\
            & adm       & 0.1685 (0.1345) & 0.3090 (0.2625) & 0.5400 (0.4575) & 0.8195 (0.7300) & 0.9625 (0.9145) \\
        \hline
        \multirow{4}{*}{(500, 500)}  
            & Exp(0.002) & 0.3195 (0.2820) & 0.7280 (0.6540) & 0.9525 (0.9085) & 0.9985 (0.9945) & 1.0000 (1.0000) \\
            & Exp(0.005) & 0.2575 (0.2095) & 0.5620 (0.4945) & 0.8575 (0.7875) & 0.9875 (0.9695) & 0.9995 (0.9975) \\
            & Exp(0.01)  & 0.1895 (0.1415) & 0.4180 (0.3360) & 0.7100 (0.6190) & 0.9410 (0.8795) & 0.9955 (0.9825) \\
            & adm       & 0.2045 (0.1670) & 0.4455 (0.3580) & 0.7415 (0.6575) & 0.9505 (0.8960) & 0.9975 (0.9900) \\
        \bottomrule
    \end{tabular}%
    }
\end{table}

Figure~\ref{Power_vs_SampleSize_prob_Alt3,4} illustrates how the power of the test based on transition probabilities improves with increasing sample size for the two Alternatives $3$ and $4$.
Under Alternative $3$, the power curves still remain separated across the censoring mechanisms as the sample size increases. Under Alternative $4$, they converge more rapidly towards one and good power of above $0.7875$ is achieved uniformly for all censoring mechanisms and sample sizes equal to or larger than $(n_1,n_2)=(300,300)$.
The unbalanced design $(n_1,n_2)=(250,450)$ generally produces rejection probabilities close to those obtained for the balanced design $(n_1,n_2)=(300,300)$.
The direction of the difference is not uniform. For example, under Alternative $3$ the rejection probability is slightly lower for $(n_1,n_2)=(250,450)$ under Exp$(0.002)$ censoring, with values $0.8165$ and $0.8315$, but slightly higher under Exp$(0.01)$ censoring, with values $0.5090$ and $0.4870$. Thus, the procedure appears stable under the considered degree of imbalance, but the additional observations in the larger group do not produce a uniform gain in power.

\begin{figure}
    \centering
    \includegraphics[width=0.78\textwidth]{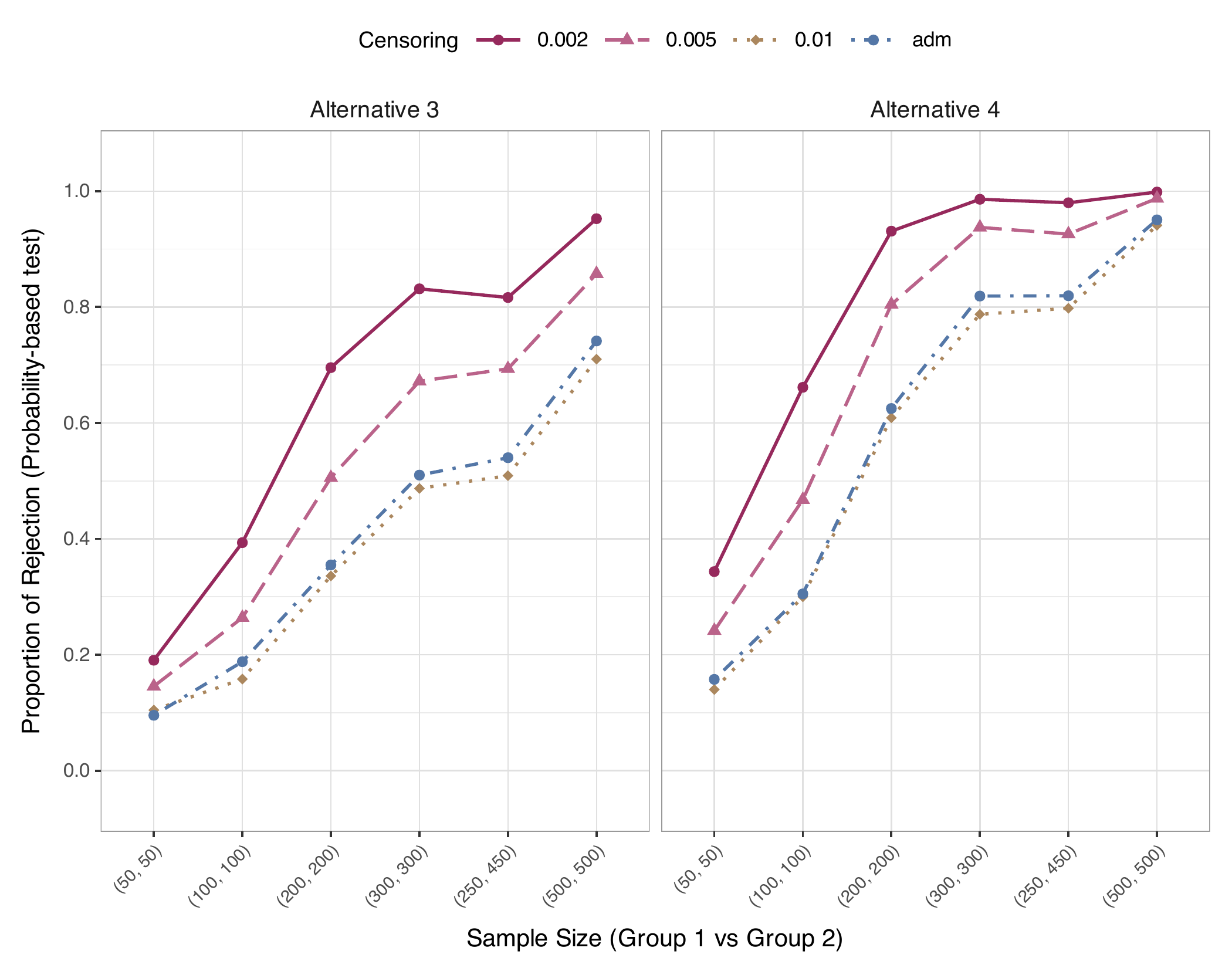}
    \caption{
    Empirical rejection probabilities of the similarity test based on transition probabilities under alternative $3$ (left) and alternative $4$ (right), in dependence of the group sample sizes. The lines represent the power for different censoring mechanisms.}
    \label{Power_vs_SampleSize_prob_Alt3,4}
\end{figure}

\section{Asymptotic Properties of the Similarity Test}
\label{appendix}

The following result shows that the test defined by Algorithm~\ref{alg1} has asymptotic level $\alpha$ and is consistent for the hypotheses in \eqref{Eq:Hypotheses} under administrative and exponential random right censoring. Note that the result is stated with respect to the true quantile $q^*_\alpha$ of the bootstrap test statistic since the estimation error of $\hat{q}^*_\alpha$ can be made arbitrarily small by increasing the number $B$ of bootstrap iterations. 

\begin{theorem} \label{Th:BootstrapLInfty}
Let $X^{(1)}$ and $X^{(2)}$ be two independent competing risks
models with transition intensities
$\alpha_{0j}^{(\ell)}>0$, $j=1,\ldots,k$ and $\ell=1,2$.
Assume that the observations are subject either to

\begin{enumerate}
    \item[(i)] administrative censoring at a fixed time $T>0$, or
    \item[(ii)] independent exponential random right censoring, where
    \begin{equation*}
        C_i^{(\ell)}
        \overset{\mathrm{i.i.d.}}{\sim}
        \operatorname{Exp}\bigl(\psi^{(\ell)}\bigr),
        \qquad \psi^{(\ell)}>0,
    \end{equation*}
    and the censoring times are independent of the corresponding
    competing risks processes.
\end{enumerate} Assume that it holds for $n := n_1 + n_2$ that
    \begin{equation*}
        \lim_{n_1,n_2 \to \infty} \frac{n}{n_1} = c \in (1, \infty).
    \end{equation*}
	 \begin{itemize}
		\item[(1)] Let $F_{\mathcal{Z}}$ denote the distribution function of the random variable  $\mathcal{Z}$ defined in equation \eqref{Eq:RandomVariableZ} below  and assume that  
        $F_{\mathcal{Z}}$ 
        is continuous at its $\alpha$-quantile $q_{\alpha,\mathcal{Z}}$.
		If the null hypothesis $H_{0}$ in \eqref{Eq:Hypotheses} holds and if $q_{\alpha,\mathcal{Z}} < 0$, then,
		\begin{equation*}
			\limsup_{n_1,n_2 \to \infty} \mathbb{P}(\hat{d}_\infty < q_{\alpha}^*) \leq \alpha.
		\end{equation*}
        \item[(2)] If the null hypothesis $H_{0}$ in \eqref{Eq:Hypotheses} holds and if the set
		$\mathcal{E}$ defined in \eqref{Def:Epsilon}
		consists of one point, it follows for any $\alpha \in (0,0.5)$ that
		\begin{equation*}
			\lim_{n_1,n_2 \to \infty} \mathbb{P}(\hat{d}_\infty < q_{\alpha}^*) =
			\begin{cases}
				0 & \text{if } d_\infty > \varepsilon_\infty,\\
				\alpha & \text{if } d_\infty = \varepsilon_\infty.
			\end{cases}
		\end{equation*}
		\item[(3)] If the alternative hypothesis $H_{1}$ in \eqref{Eq:Hypotheses} holds, it follows for any $\alpha \in (0,0.5)$ that
		\begin{equation*}
			\lim_{n_1,n_2 \to \infty} \mathbb{P}(\hat{d}_\infty < q_{\alpha}^*) = 1.
		\end{equation*}
	\end{itemize}
\end{theorem}

\begin{proof}
    We start by stating the full proof for the case of administrative censoring.
    By Theorem $6.1$ in \citet{Albert}, the vector
    \begin{equation} \label{appendix:NormalDistribution}
        \sqrt{n} ((\hat{\alpha}^{(1)}-\alpha^{(1)})^T,(\hat{\alpha}^{(2)}-\alpha^{(2)})^T)^T
    \end{equation}
    converges weakly to a multivariate normal distribution with mean $0$ and diagonal covariance matrix.
 Define the set $\mathcal{M} := [0, \tau] \times \{1, \dots, k\}$ and let $\ell^\infty ({\cal M})$ denote the set of bounded functions $g: {\cal M} \to \mathbb{R}$. The function mapping the transition intensities to the respective transition probability function is, based on \eqref{Eq:ProbabilityFromIntensity}, given by
    \begin{equation*} 
	P :
	\left\{
	\begin{alignedat}{2}
		(0, \infty)^k &\to \ell^\infty(\mathcal{M})\\
		\alpha &\mapsto  P(\alpha) : 
        \begin{cases}  
        \mathcal{M}  &\to  [0,1] \\
        (t,j) &\mapsto - \frac{(-1+ \exp (-{\sum_{i=1}^k \alpha_{0i} t})) \alpha_{0j}}{\sum_{i=1}^k \alpha_{0i}}
              \end{cases}
	\end{alignedat}
	\right. ~. 
\end{equation*}    
As this function is differentiable, we obtain  
the following weak convergence by an application of the functional delta method of \citet[p. 297]{vanDerVaart}.
\begin{equation} \label{Eq:GaussianProcess}
\begin{split}
    &\left\{ \sqrt{n} \left( \hat{P}_{0j}^{(1)}(t) - P_{0j}^{(1)}(t)
    - \bigl(\hat{P}_{0j}^{(2)}(t) - P_{0j}^{(2)}(t)\bigr) \right) \right\}_{(t,j) \in \mathcal{M}} \\
    &\hspace{7em}\leadsto \left\{ G_{0j} (t)\right\}_{(t,j) \in \mathcal{M}} .
\end{split}
\end{equation}
in $\ell^\infty(\mathcal{M})$ where $\left\{ G_{0j} (t)\right\}_{(t,j) \in \mathcal{M}}$ is a centered Gaussian process. 
This convergence statement is equivalent to equation $(A.7)$ in \citet{DetteMoellenhoff} and by applying the functional delta method for directionally Hadamard differentiable maps of \citet{Carcamo}, we get
\begin{equation} \label{Eq:RandomVariableZ}
\begin{split}
    \sqrt{n} \Bigg(&\max_{(t,j) \in \mathcal{M}}
    \left\vert \hat{P}_{0j}^{(1)}(t)-\hat{P}_{0j}^{(2)}(t) \right\vert \\
    &-\max_{(t,j) \in \mathcal{M}}
    \left\vert P_{0j}^{(1)}(t)-P_{0j}^{(2)}(t) \right\vert\Bigg) \\
    &\leadsto \mathcal{Z}:=\max\left\{
    \max_{(t,j) \in \mathcal{E}^+}G_{0j}(t),
    \max_{(t,j) \in \mathcal{E}^-}-G_{0j}(t)\right\}.
\end{split}
\end{equation}
where
\begin{equation*}
    \mathcal{E}^{\pm} := \left\{ (u,i) \in \mathcal{M} \: \Big\vert \:
    P_{0i}^{(1)}(u) - P_{0i}^{(2)}(u) = \pm d_\infty \right\}.
\end{equation*}
This statement is analogous to the statement of Theorem $3$ in \citet{DetteMoellenhoff}.

Recall that $\hat{\alpha}^{*(1)}$ and $\hat{\alpha}^{*(2)}$ are the MLEs of the bootstrap transition intensities defined in Step $3.2$ of Algorithm~\ref{alg1}, $\hat{\hat{\alpha}}^{(1)}$ and $\hat{\hat{\alpha}}^{(2)}$ are the possibly constrained estimators and $\Tilde{\alpha}^{(1)}$ and $\Tilde{\alpha}^{(2)}$ are the constrained estimators.
Note that $\Tilde{\alpha}^{(1)}$ and $\Tilde{\alpha}^{(2)}$ are MLEs computed by optimizing the log-likelihood function in \eqref{Eq:LikelihoodTypeOneCensoring} over the closed parameter space
\begin{equation*} 
    \Theta = \Big \{ (\alpha^{(1)},\alpha^{(2)}) \in \mathbb{R}_{\geq 0}^{2k} \: \big\vert \: \max_{j=1}^k \max_{t \in [0,\tau]} \big \vert P^{(1)}_{0j}(t) - P^{(2)}_{0j}(t) \big \vert = \varepsilon_\infty \Big \}.
\end{equation*}
By Theorem $2$ in \citet{Wald}, it therefore follows under regularity conditions that the estimators are consistent if $(\alpha^{(1)},\alpha^{(2)}) \in \Theta$. It follows as in \citet{DetteMoellenhoff} that under the null hypothesis 
\begin{equation} \label{Eq:ConsistencyConstrainedEstimator}
    \hat{\hat{\alpha}}^{(\ell)} \overset{\mathbb{P}}{\longrightarrow} \alpha^{(\ell)}  ~~~~(\ell =1,2) .
\end{equation}
We further have to show the weak convergence in probability for the bootstrap statistic $\hat{d}_{\infty}^*$ (for a definition of weak convergence in probability see \citet[p. 326ff.]{vanDerVaart}). For this purpose, we note that the bootstrap data is sampled from a different distribution for each sample size $n$ and that it can be shown using the consistency of the constrained estimator in \eqref{Eq:ConsistencyConstrainedEstimator} that a conditional version of the Lindeberg-Feller conditions holds; see \citet[p. 96ff.]{Lange}. An application of the corresponding conditional version of the Lindeberg-Feller theorem in \citet[p. 96ff.]{Lange} shows that the vector
\begin{equation*}
    \sqrt{n} \left( (\hat{\alpha}^{*(1)} - \hat{\hat{\alpha}}^{(1)})^T, ( \hat{\alpha}^{*(2)} - \hat{\hat{\alpha}}^{(2)} )^T \right)^T
\end{equation*}
converges weakly conditionally, in probability, on the combined sample from groups $1$ and $2$ to a multivariate normal distribution with mean $0$ and diagonal covariance matrix.
Similar to the proof of Theorem $4$ in \citet{DetteMoellenhoff}, we can linearize the function
\begin{equation} \label{Eq:AppendixBootstrapProbabilities}
    t \mapsto \sqrt{n} \left( \hat{P}^{*(1)}_{0j}(t) - \hat{\hat{P}}^{(1)}_{0j}(t)- \left( \hat{P}^{*(2)}_{0j}(t) - \hat{\hat{P}}^{(2)}_{0j}(t) \right) \right)
\end{equation}
by an application of Taylor's theorem. With the consistency of the constrained estimator given in \eqref{Eq:ConsistencyConstrainedEstimator}, it follows that the function in \eqref{Eq:AppendixBootstrapProbabilities} converges weakly conditionally on the sample in probability to the Gaussian process $\left\{ G_{0j} (t)\right\}$ defined in \eqref{Eq:GaussianProcess}. This result is analogous to equation $(A.25)$ in \citet{DetteMoellenhoff}, and we are now in the position to prove the three parts of Theorem \ref{Th:BootstrapLInfty}.

\begin{itemize}
    \item[(1)] The same arguments used to infer $(A.32)$ in \citet{DetteMoellenhoff} show that $\sqrt{n}(\hat{d}_\infty^* - \hat{\hat{d}}_\infty)$ is bounded from above by $\mathcal{Z}_n^* + o_{\mathbb{P}}(1)$, where $\mathcal{Z}_n^*$ is a bootstrap statistic converging weakly conditionally on the sample in probability to $\mathcal{Z}$ and $o_{\mathbb{P}}(1)$ is a term converging to $0$ in probability for $n \to \infty$.
From there, the proof of $(1)$ is analogous to the proof of $(4.9)$ in Theorem $4$ in \citet{DetteMoellenhoff}.
\item[(2)] If $\mathcal{E} = \{ (t_0,j_0) \}$, then it holds that
\begin{equation*}
    \sqrt{n}(\hat{d}_\infty^* - \hat{\hat{d}}_\infty) \leadsto G_{0j_0}(t_0)
\end{equation*}
conditionally on the sample in probability. This is equivalent to statement $(A.30)$ in \citet{DetteMoellenhoff}, and the proof follows as the proof of $(4.10)$ in that article.
\item[(3)] This result follows by the same arguments as given in the derivation of $(3.18)$ in \citet{DetteMoellenhoff}.
\end{itemize}

For the case of exponential random right censoring note that by Theorem $6.1.2$ in \citet{AndersenCountingProcesses} the vector in \eqref{appendix:NormalDistribution} converges weakly to a multivariate normal distribution with mean $0$ and diagonal covariance matrix as well. Note that this covariance matrix depends on the censoring parameters $\psi^{(1)}$ and $\psi^{(2)}$. By the delta method, the estimated transition probability curves converge weakly to a Gaussian process with the corresponding covariance structure. The subsequent continuous mapping and supremum arguments are unchanged. The bootstrap remains valid because the bootstrap sample has the same model structure as the original sample with $(\alpha^{(\ell)}, \psi^{(\ell)})$ replaced by $(\hat{\hat{\alpha}}^{(\ell)}, \hat{\psi}^{(\ell)})$. By the consistency in \eqref{Eq:ConsistencyConstrainedEstimator} and $\hat{\psi}^{(\ell)} \overset{\mathbb{P}}{\longrightarrow} \psi^{(\ell)}$, the conditional covariance matrix of the bootstrap MLEs converges in probability to the corresponding limiting covariance matrix. Then, the result follows analogous to the administrative censoring case.
 
\end{proof}

\end{document}